\documentclass[12pt, fleqn]{article}
\usepackage{amsmath,amsfonts}
\begin{document}
\title{Towards a Covariant Theory of Gravitation\footnote{Published in \emph{Abstracts of Contributed Papers} for the 9th International Conference on General Relativity and Gravitation, Jena, GDR, 1980 (This rendition of the paper is an improved English translation from the Russian original and varies slightly from the published paper.)}
}
\author{Alexander Poltorak\footnote{Current address: General Patent Corporation, Montebello Park, 75 Montebello Road, Suffern, NY 10901-3740 USA. E-mail: apoltorak@gpci.com.}\\
\\\normalsize Kuban University\\\normalsize Department of Theoretical Physics\\\normalsize Krasnodar, USSR}
\date{1980}
\maketitle

\begin{abstract}
In the following GR9 (1980) Conference abstract, we briefly considered a covariant reformulation of the GR from three points of view: geometrodynamics, Lagrange-Euler field theory, and gauge field theory.  From a perspective of geometrodynamics, a definition of a reference frame as a differential manifold with an affine connection results in separation of the respective contributions of inertial and gravitational fields represented by the affine connection and the nonmetricity tensor within the Levi-Civita connection of GR.  Resulting decomposition of Riemannian, Ricci and Einstein curvature tensors into affine and nonmetric parts allows recasting of Einstein's field equations in a form invariant with respect to the choice of a reference frame, wherein the gravity is described by nonmetricity of space-time.  A covariant Lagrangian is proposed leading to the same field equation.  All three approaches ultimately lead to the same fully covariant theory of gravitation with a covariant tensor of energy-momentum of the gravitational field and differential and integral conservation laws.  The role of reference frames, as distinguished from coordinate systems, is discussed.
\end{abstract}
\medskip
\normalsize
General Relativity (GR), as it is generally known, is not a consistently covariant theory, because covariant field equations coexist with non-covariant conservation laws.  In our opinion, this inconsistency arises from the fact that a non-covariant Riemannian (Cristoffel) connection describes a gravitational field in a chosen frame of reference, but not the field itself.  Consequently, we are faced with the problem of separating the respective contributions of a gravitational field and a field of inertia into the Riemannian geometry.  We will attempt to resolve this problem based on the assumption that a frame of reference is represented by a differential manifold wherein an affine connection serves as a potential of inertia field generated by this frame of reference.

\indent \emph{Theorem}.  Let $M$ be a Riemannian manifold with the metric $g$; $BM$ be a set of all differentiable vector fields on $M$; \(
\nabla\) be the Levi-Civita connection (a Riemannian connection without torsion) on \emph{M}; \(
{\bar \nabla }\) be the affine connection; and \emph{R} and \({\bar R}\) be the curvature tensors of the connection \(\nabla\) and \(
{\bar \nabla }\) respectively.  Then,
\begin{enumerate}
\item There exists one and only one tensor \emph{S}: $BM \times BM \to BM{\rm  }$ (written as $S(X,{\rm  }Y) = S_x Y{\rm  }){\rm  }
$
 that satisfies the conditions $\forall x,y,z \in BM$

$$
\overline \nabla  _z g\left( {X,Y} \right) = g\left( {S_z X,Y} \right) + g\left( {X,S_z Y} \right)
$$
$$
S_x Y{\rm  } = {\rm  }S_y X
$$

\item $\nabla  = \overline \nabla   + S$

\item \(
R = \overline R  + \hat R\), where

\[
\hat R\left( {X,Y} \right) = \left[ {\overline \nabla  _x ,S_y } \right] + \left[ {S_x ,\overline \nabla  _y } \right] + \left[ {S_x ,S_y } \right] + S\left[ {X,Y} \right]
\]

\end{enumerate}

Suppose, \emph{x} is a coordinate system on $M$; $X_\mu   = {d \over {d_\mu  }}$, $g\left( {X_\mu  ,X_\nu  } \right) = g_{\mu \nu }$, $Sx_\mu  X_\nu   = S_{\mu \nu }^\lambda  X_\lambda
$, $R\left( {X_\lambda  ,X_\mu  } \right)X_\nu   = R_{\lambda \mu \nu }^\varepsilon  X_\varepsilon$, then we have in the coordinate system
\emph{x}

\[
g_{\tau \sigma } S_{\mu \nu }^\tau   = \frac{1}
{2}\left( {\overline \nabla  _\mu  g_{\nu \sigma }  + \overline \nabla  _\nu  g_{\mu \sigma }  - \overline \nabla  _\sigma  g_{\mu \nu } } \right)
\]

\[
\hat R_{\lambda \mu \nu }^\varepsilon   = \overline \nabla  _\lambda  S_{\mu \nu }^\varepsilon   - \overline \nabla  _\mu  S_{\lambda \nu }^\varepsilon   + S_{\lambda \tau }^\varepsilon  S_{\mu \nu }^\tau   - S_{\lambda \mu }^\varepsilon  S_{\nu \tau }^\tau
\]Tensor \(S_{\mu \nu }^\lambda\) is called the tensor of nonmetricity.

\medskip
\emph{Corollary:} Let \(G = Sp * R *\) be an Einstein tensor, then \(
G = \bar G + \hat G\) where \(\bar G = Sp * \bar R *\) and \(\hat G = Sp * \hat R *\).

This theorem allows us to rewrite the equation for the geodesic lines \[
\nabla  \times X = 0
\]

as
\[\bar \nabla  \times X =  - S \times X\] which means: test particles move in a space of affine connection under the influence of a force field which we shall identify with the gravitational field.  According to this, Einstein's field equation
\[G = 8\pi T\]

(see notation in ref.[1]) takes the form

\[\hat G = 8\pi T - \bar G\] (Tensor \({\bar G}\) may be considered as the energy-momentum tensor of the inertia field).  This equation is the gravitational field equation in an arbitrary frame of reference and differs from Einstein's equation insofar as its solution requires defining the inertia field, i.e. affine connection, along with the energy-momentum tensor of matter \emph{T}.

The laws of conservation for the matter and gravitational field take the form
\[
\bar \nabla  \cdot \left( {T + t} \right) = 0
\]
where \emph{t} is the covariant energy-momentum tensor of the gravitational field.  The existence of integral laws of conservation depends on the symmetry of the inertia field only.  So, in a frame of reference generating symmetric space, there always exists a decuplet of conservative integral values
\[
P = \int \xi  ^a  \cdot \left( {T + t} \right) \cdot d\Sigma
\] (\(\xi ^a\) are Killing's vectors) giving the regular representation of the Lie algebra of the isometric group of the frame of reference space.

It is of interest to consider the problem from the standpoint of the Lagrangian field theory.  Einstein's field equations arise from the variational principle in which a scalar curvature of the Riemannian space is used as a Lagrangian.  General Relativity is a local theory.  From the standpoint of the global geometry, a differentiable field has two local representations in the coordinate system x:
\begin{enumerate}
\item \(R = g^{\mu \sigma } \left( {\partial _\mu  \Gamma _{\nu \sigma }^\nu   - \partial _\nu  \Gamma _{\mu \sigma }^\nu   + \Gamma _{\mu \lambda }^\nu  \Gamma _{\nu \sigma }^\lambda   - \Gamma _{\nu \lambda }^\nu  \Gamma _{\mu \sigma }^\lambda  } \right)
\)
\item \(R = g^{\mu \sigma } \left( {\bar \nabla _\mu  D_{\nu \sigma }^\nu   - \bar \nabla _\nu  D_{\mu \sigma }^\nu   + D_{\mu \lambda }^\nu  D_{\nu \sigma }^\lambda   - D_{\nu \lambda }^\nu  D_{\mu \sigma }^\lambda  } \right)
\)
\end{enumerate}
\[
D_{\mu \nu }^\sigma   = \Gamma _{\mu \nu }^\sigma   - \bar \Gamma _{\mu \nu }^\sigma\] where \(\bar \Gamma _{\mu \nu }^\sigma\) are the Cristoffel connection coefficients, \(\bar \Gamma _{\mu \nu }^\sigma\)
are the absolute parallelism connection coefficients, and \(\bar \nabla _\mu\) is the covariant derivative with respect to \(\bar \Gamma _{\mu \nu }^\sigma\).

In General Relativity, the first representation \(L = R\left( {g_{\mu \nu } ,\Gamma _{\mu \nu }^\lambda  ,\partial _\varepsilon  \Gamma _{\mu \nu }^\lambda  } \right)
\) is used where only the first variable \({g_{\mu \nu } }\) is a tensor.  Consequently, the field equation resulting from variation of the \({g_{\mu \nu } }\) turns out covariant, but the energy-momentum tensor of a gravitational field which contains \(
\Gamma _{\mu \nu }^\lambda\) turns out not to be a tensor.  Generally, variables ${g_{\mu \nu } }$ and $\Gamma _{\mu \nu }^\lambda  $ are objects of different algebraic natures and do not form a configurational space by any reasonable definition of this concept.  So a gravitational field with the Lagrangian $L = R$
in the first representation is not a Lagrangian Dynamic system in a rigorous sense of the word, and therefore the choice of the first representation is not appropriate.

The second representation used in bimetrism [2] does not lead to the abovementioned difficulties because all the variables in the Lagrangian $L = R\left( {g_{\mu \nu } ,D_{\mu \nu }^\sigma  ,\bar \nabla _\varepsilon  D_{\mu \nu }^\sigma  } \right)$
are tensors.  This is why we find the covariant field equation in bimetrism side by side with covariant laws of conservation.

Let us return now to General Relativity.  The connection coefficients form the affine tensor, so that if we limit ourselves to the affine subgroup $G_\alpha  $
of the general coordinates transformation group $G_\infty  $, we shall be able to construct the correct theory based on the Lagrangian $L = R\left( {g_{\mu \nu } ,\partial _\sigma  g_{\mu \nu } ,\partial _\varepsilon  \partial _\sigma  g_{\mu \nu } } \right)$.

Let us localize the parameters of group $G_\alpha  $ making them the local functions and in that way transform $G_\alpha  $ into gauge group $G_{\alpha \infty } $.  The Lagrangian \emph{R} is a scalar, but to assure the gauge invariance of the theory, it is necessary and sufficient to replace in the Lagrangian $\partial _\mu   \to \delta _\mu ^\lambda  \bar \nabla _\nu   = \delta _\mu ^\lambda  \partial _\nu   - \bar \Gamma _{\mu \nu }^\lambda  $, i.e. to introduce the interaction with the gauge inertia field, the potential of which is the affine connection ${\bar \nabla }$.

The Lagrangian $L = L\left( {g_{\mu \nu } ,\bar \nabla _\sigma  g_{\mu \nu } ,\bar \nabla _\varepsilon  \bar \nabla _\sigma  g_{\mu \nu } } \right)$
derived by replacement of partial derivatives by covariant ones turns out to be equal to the non-metric part of scalar curvature $L = R = S_p  * \hat R *  \cdot g$.  If we add the Lagrangian of interaction with gauge inertia field $L_{in}  = Sp * \bar R *  \cdot g$ and the Lagrangian of matter to the field Lagrangian, we shall obtain the Lagrange equation for the gravitational field $\hat G = 8\pi T - \bar G$, derived above from the geometrical consideration.

Nonlocalization of gravitational energy is such a serious problem that attempts have been made to justify this fact by means of the equivalence principle [1].  This approach is a result of confusing the "frame of reference" with a coordinate system.  I subscribe to the other point of view that coordinates have no physical meaning and non-inertial frame define a space of affine connection with curvature [3].

Adhering to this position, I suggest two ways of solving the problem in the context of the covariant theory of gravitation.  The first geometric approach, based on the idea of the separation of the respective contributions of gravitation and inertia in the Riemannian geometry, is realized by means of splitting the Riemannian connection and its curvature tensor into affine and nonmetric parts, due to which Einstein's equation takes the form suitable for the description of gravitation in an arbitrary frame of reference.  Following this approach, it becomes possible to obtain covariant differential and integral laws of conservation for gravitational field.

The second field-theory approach is based on the fact that for the gauge invariance of the field theory, the invariance of the Lagrangian is necessary but not sufficient.  To ensure gauge invariance, the Lagrangian must be a function of only covariant (tensor) arguments.  This second condition is not satisfied in General Relativity.  Introducing the interaction with a gauge field we ensure covariance of the theory which coincides with the theory that was obtained through the geometric approach outlined above.

In conclusion, I should like to note that it is attractive to interpret the inertia field as a gauge field because it makes possible the application of the methods of gauge theory for the description of the non-inertial frames of reference.

\bigskip
\large{References:}
\normalsize
\begin{enumerate}
\item C. Misner, K. Thorne, J. Wheeler.  Gravitation. Freeman, San Francisco, 1973.
\item N. Rosen.  Phys. Rev., 57, 147, 1940.
\item V. Rodichev. Theory of Gravitation in Orthogonal Frame. Nauka, Moscow, 1974. (In Russ.)
\end{enumerate}
\end{document}